\documentclass[
twocolumn,
print,
showpacs,
preprintnumbers,
 amsmath,amssymb,
 prd,
]{revtex4-1}
\usepackage{cases}
\usepackage{footmisc}
\usepackage{graphicx}
\usepackage{dcolumn,color}
\usepackage{bm}

\begin{document}

\title{Linearization of thick $K$-branes}

\author{Yuan Zhong\footnote{zhongy2009@lzu.edu.cn},
        Yu-Xiao Liu\footnote{liuyx@lzu.edu.cn, corresponding author.}
        }
\affiliation{
    Institute of Theoretical Physics, Lanzhou University,
           Lanzhou 730000, People's Republic of China}

\begin{abstract}
We study the linearization of a class of thick K-branes, namely, four-dimensional domain walls generated by a scalar field with particular nonstandard kinetic terms. The master equations for linear perturbations are derived from the point of view of both dynamical equations and quadratic action. The spectra of the canonical normal modes are studied using supersymmetric quantum mechanics. Our results indicate that the scalar perturbation is nonlocalizable in general. Conditions for stable $K$-branes are also found.
\end{abstract}

\maketitle


\section{Introduction}
The $K$-field, namely, a scalar field with nonstandard kinetic terms, was firstly introduced to establish a new mechanism of inflation in cosmology~\cite{Armendariz-PiconDamourMukhanov1999,GarrigaMukhanov1999,Armendariz-PiconMukhanovSteinhardt2001}.
However, with the development of brane-world scenarios~\cite{RubakovShaposhnikov1983,RandallSundrum1999,RandallSundrum1999a,
DeWolfeFreedmanGubserKarch2000,Gremm2000,
Gremm2000a,CsakiErlichHollowoodShirman2000,Giovannini2001a,Giovannini2003,
Giovannini2002} (see Refs.~\cite{Rubakov2001,Csaki2004,Shifman2010,DzhunushalievFolomeevMinamitsuji2010} for reviews), the $K$-field was applied frequently in brane-world models. For example, the $K$-field was employed to model a smooth version of the negative 3-brane~\cite{KoleyKar2005a} that appears in the Randall-Sundrum-I brane-world scenario~\cite{RandallSundrum1999}; to stabilize the distance between thin branes~\cite{NunesPeloso2005,Olechowski2008,MaitySenGuptaSur2009} via the Goldberger-Wise mechanism~\cite{GoldbergerWise1999,LesgourguesSorbo2004};
to offer a new mechanism of field localization~\cite{AdamGrandiSanchez-GuillenWereszczynski2008}; or to construct new brane-world solutions~\cite{BazeiaBritoNascimento2003,KoleyKar2005b,AdamGrandiKlimasSanchez-GuillenWereszczynski2008,BazeiaGomesLosanoMenezes2009,LiuZhongYang2010,CastroMeza2013,BazeiaDantasGomesLosanoMenezes2011,BazeiaLobaoMenezes2012}, and so on.

One of the important issues in brane-world models is the linearization of the system. For one thing, linearization is a key procedure for the study of the stability of the classical brane solution~\cite{Giovannini2001a,KobayashiKoyamaSoda2002}.
For the other, to reproduce the four-dimensional Newtonian potential and its short distance modification, we need also to study the linear structure of the system~\cite{CsakiErlichHollowoodShirman2000,Giovannini2003}.
The linearization of the standard thick branes (namely, models with standard bulk scalar field) was extensively studied in Refs.~\cite{Giovannini2001a,Giovannini2003}.

As to the thick $K$-branes, tensor perturbation and the localization of gravity were discussed in Ref.~\cite{BazeiaGomesLosanoMenezes2009}. The study indicates that the introduction of the $K$-field does not affect the pattern of the tensor perturbation. The stability of the domain wall solution under only matter perturbation was discussed in Refs.~\cite{AdamGrandiSanchez-GuillenWereszczynski2008,AdamGrandiKlimasSanchez-GuillenWereszczynski2008}. But the complete discussion should contain both matter and metric perturbations. However, the scalar part of the metric perturbations usually couples with the matter perturbation, that renders the discussion a nontrivial work. To our knowledge, a systematical discussion on the linearization of the thick $K$-brane is still lacking in the literature.

Therefore, in this paper, we try to give a general and systematical discussion on the linearization of a class of $K$-brane models. We will study the linearization of our model, both by linearizing the dynamical equations to first order, and by perturbing the action to the second order. We consider both approaches, because as stated in Ref.~\cite{Giovannini2003}, they are only partly equivalent. More importantly, the normal modes of the perturbations can be found only from the action approach. Our aim is to figure out whether the modification in the matter Lagrangian affects the localization of the scalar zero mode, and to what extent, a classical $K$-brane solution is stable.

In the next section, we present the setup of the model and derive the background field equations. In Sec.~\ref{sec3}, we linearize the dynamical equations. The master equations are obtained by using the scalar, tensor, and vector (STV) decomposition of the metric perturbations. The issue of gauge invariance of the master equations is also discussed briefly in this section. Then in Sec. \ref{sec4}, we reconsider the linearization of our model from the point view of quadratic action of perturbations. The normal mode of each type of perturbations is found. In the end, we give a brief summary on our results.

\section{$K$-brane model and background equations}
\label{sec2}
In the present paper, we study a model with the following action:
\begin{eqnarray}
S=\int d^5x \sqrt{-g}\left(\frac{1}{2\kappa_5^2}R+\mathcal{L}(\phi,X)\right),
\end{eqnarray}
where $\kappa^2_5= 8\pi G_5$ is the gravitational coupling constant and $G_5$ is the five-dimensional Newtonian constant. $X\equiv-\frac12g^{MN}\nabla_M\phi\nabla_N\phi$ represents the kinetic term of the
scalar field; the model with $\mathcal{L}=X-V(\phi)$ is referred to as the standard model. In this paper, we always use $\mu, \nu=0,1,2,3$ to denote the indices of brane coordinates and use $M, N=0,1,2,3,5$ to represent the indices of bulk coordinates.

The Einstein equations are
\begin{eqnarray}
\label{eq1}
G_{MN}\equiv R_{MN}- \frac{1}{2}g_{MN}R = \kappa _5^2{T_{MN}},
\end{eqnarray}
where
\begin{eqnarray}
\label{eq2}
T_{MN}={g_{MN}}\mathcal{L} + \mathcal{L}_X\nabla_M\phi \nabla_N\phi
\end{eqnarray}
is the energy-momentum tensor. In this paper, $\mathcal{L}_X\equiv\partial\mathcal{L}/\partial X$ and $\mathcal{L}_{\phi}\equiv\partial\mathcal{L}/\partial\phi$ and so on.

The metric for a domain wall brane is assumed to be
\begin{eqnarray}
\label{metricconf}
g_{MN}=a^2(r)\eta_{MN},
\end{eqnarray}
from which, we immediately obtain the Einstein equations:
\begin{subequations}
\label{Einstin}
\begin{eqnarray}
\label{Einstin1}
6{\left( {\frac{{a'}}{a}} \right)^2}& =& \kappa _5^2{a^2}{\cal L} + \kappa _5^2{{\cal L}_X}\phi {'^2},\\
\label{Einstin2}
3\frac{{a''}}{a}& =& \kappa _5^2{a^2}{\cal L},
\end{eqnarray}
\end{subequations}
where primes denote the derivatives with respect to the extra-dimensional coordinate $x^5=r$.

The equation of motion for the scalar field is
\begin{eqnarray}
\label{EOM}
\phi'\mathcal{L}_X'+\mathcal{L}_X\left(\phi''+3\frac{a'}{a}\phi'\right)
=-a^2\mathcal{L}_\phi.
\end{eqnarray}
This equation can be derived from the Einstein equations as a natural result of the Bianchi identity $\nabla^N G_{MN}=0$.

\section{Linearization of the field equations}
\label{sec3}
Consider small perturbations around an arbitrary background solution $\{\bar{g}_{MN}(r), \bar{\phi}(r)\}$, so that the perturbed fields are given by
\begin{eqnarray}
\label{3.1}
 \phi&=&\bar{\phi}(r)+\delta\phi(x^P),\\
g_{MN}&=&\bar{g}_{MN}(r)+\delta{g}_{MN}(x^P).
\end{eqnarray}
It is more convenient to define $\delta g_{MN}\equiv a^2h_{MN}$. Up to the first order, the orthogonal relation $g_{MP}g^{PN}=\delta_M^N$ implies $\delta g^{MN}\equiv -g^{MP}g^{NQ}\delta g_{PQ}=-a^{-2}h^{MN}$. The indices of $h_{MN}$ are raised or lowered by $\eta_{MN}$, consequently, $h\equiv \eta^{MN} h_{MN}$.

It is always possible to make the so-called STV decomposition (see Ref.~\cite{Weinberg2008} for a similar discussion in cosmology):
\begin{subequations}
\label{decomposition}
\begin{eqnarray}
{h_{\mu r}} &=& {\partial _\mu }F + {G_\mu },\\
{h_{\mu \nu }} &=& {\eta _{\mu \nu }}A + {\partial _\mu }{\partial _\nu }B + 2{\partial _{(\mu }}{C_{\nu )}} + {D_{\mu \nu }},
\end{eqnarray}
\end{subequations}
where $C_\mu$ and $G_\mu$ are transverse vector perturbations:
\begin{eqnarray}
\partial^\mu C_\mu=0=\partial^\mu G_\mu,
\end{eqnarray}
and $D_{\mu \nu }$ is transverse and traceless perturbation:
\begin{eqnarray}
\partial^\nu D_{\mu \nu }=0=D^\mu_\mu.
\end{eqnarray}
Here all indices are raised with $\eta^{\mu\nu}$, so that $\partial^{\mu}\equiv\eta^{\mu\nu}\partial_{\nu}$. Likewise, we denote $\partial^{P}\equiv\eta^{PQ}\partial_{Q}$ and $\square^{(5)}\equiv\eta^{MN}\partial_M\partial_N$, ${\square^{(4)}}\equiv\eta^{\mu\nu}\partial_\mu
\partial_\nu$ in our following discussions.

It is well known that, due to the general covariant principle, these linear perturbation equations are invariant under the following gauge transformations (see Ref.~\cite{Mukhanov2005} for details):
\begin{eqnarray}
\label{gaugeTrans}
\Delta h_{MN}&\equiv& \widetilde{h}_{MN}-h_{MN}\nonumber\\
&=&-2\xi_{(M,N)}
   -2\eta_{MN}\frac{a'}{a}\xi^r,
\end{eqnarray}
and
\begin{eqnarray}
\label{transphi}
\Delta \delta\phi=-\xi^r \bar{\phi}'.
\end{eqnarray}
Here, we use ``$\Delta$" to indicate the change of perturbations, and $\xi_M\equiv\eta_{MN}\xi^N$ relates to an infinitesimal transformation of the coordinate
\begin{eqnarray}
x^M\to {\tilde{x}^M}=x^M+\xi^M(x^P).
\end{eqnarray}

Since $(\delta\phi, h_{MN})$ and $(\widetilde{\delta\phi}, \widetilde{h}_{MN})$ satisfy the same equations, nonphysical perturbations exist due to our freedom in choosing $\xi^M$. We can eliminate the nonphysical freedoms by taking gauge directly~\cite{KakushadzeLangfelder2000}, for instance, we can take $\widetilde{\delta\phi}=0$, simply by asking
\begin{eqnarray}
\label{4.7}
\xi^r=\frac{ \delta\phi}{\bar{\phi}'}.
\end{eqnarray}
Likewise, we can eliminate some other perturbations by using the residual freedoms in choosing $\xi^\mu$. Some authors prefer to take gauges in the light-cone coordinates~\cite{Randjbar-DaemiShaposhnikov2002}.
However, it is difficult to eliminate all the gauge freedoms completely if we directly take gauges.

Nevertheless, with the decomposition we introduced previously, we are ready to construct gauge-invariant quantities which not only completely fix the gauge freedom, but can serve as the physical dynamic variables in the quantization procedure. This method was introduced to study cosmological perturbations~\cite{MukhanovFeldmanBrandenberger1992}, and then generalized in brane-world models~\cite{Giovannini2001a,Giovannini2002}.

Using the properties of the decomposed metric perturbations, the gauge transformation Eq.~\eqref{gaugeTrans} can be rewritten as
\begin{eqnarray}
\Delta A&=&-2\frac {a'}{a}\xi^r, \quad
\Delta h_{rr}=
   -2\xi^{r\prime}
   -2\frac{a'}{a}\xi^r,\nonumber\\
\Delta B&=&-2\zeta,\quad
\Delta F=-\xi^r-\zeta',\nonumber\\
\Delta C_{\mu}&=&-\xi^{\perp}_{\mu},\quad
\Delta G_\mu=-\xi_\mu^{\perp\prime},\nonumber\\
\Delta D_{\mu\nu}&=&0.
\end{eqnarray}
Here, we applied the decomposition $\xi^\mu=\partial^\mu\zeta+\xi^{\perp\mu}$ such that $\partial_\mu\xi^{\perp\mu}=0$.

Defining $\psi  = F - \frac{1}{2}B'$ and ${v_\mu } = {G_\mu } - C_\mu'$, we conclude that $\Delta\psi=-\xi^r$, while ${v_\mu }, D_{\mu\nu}$ and the following scalar quantities are invariant under gauge transformations:
\begin{eqnarray}
\label{Ginvariant1}
\Xi&\equiv& h_{rr}-2\frac1a \left(a\psi\right)',\\
\label{Ginvariant2}
\Psi&\equiv& A-2\frac{a'}{a}\psi,\\
\label{Ginvariant3}
\Phi&\equiv&\delta\phi-\phi'\psi.
\end{eqnarray}

Following a similar discussion in Ref.~\cite{Giovannini2001a}, one would obtain the master equations for the vector
\begin{eqnarray}
\label{vc1}
  {\square ^{(4)}}{v_\mu } &=& 0,\\
\label{vc2}
  3\frac{{a'}}{a}{\partial _{(\mu }}{v_{\nu )}} + {\partial _{(\mu }}v{'_{\nu )}} &=& 0,
\end{eqnarray}
and tensor perturbation:
\begin{eqnarray}
  \label{tens}
\square ^{(4)}D_{\mu \nu }+ D_{\mu \nu }'' + 3\frac{{a'}}{a}D_{\mu \nu }' &=& 0.
\end{eqnarray}

Because of the decoupling of different perturbation types, the modification of the matter Lagrangian does not affect the vector and tensor parts. So, here we just summarize some of the main conclusions of Ref.~\cite{Giovannini2001a}:
\begin{enumerate}
  \item The tensor and vector perturbations do not destroy the stability of brane solutions. The vector perturbation supports only zero mode, while the tensor perturbation usually permits both zero mode and a series of massive modes.
  \item If we demand a finite four-dimensional Planck mass and the reproduction of the four-dimensional Newtonian gravity, the zero mode of tensor perturbation must be localized on the brane. As a cost, the vector zero mode cannot be localized due to Eq.~\eqref{vc2}.
\end{enumerate}

Similarly, we can express the scalar perturbation equations in terms of the gauge-invariant quantities:
\begin{eqnarray}
\label{18}
  &&- \Psi - \frac{1}{2}{\Xi} = 0,  \\&&
\label{19}
  \frac{3}{2}\frac{{a'}}{a}{\Xi} - \frac{3}{2}\Psi' = \kappa _5^2{\mathcal{L}_{X}}\phi '\Phi,\\&&
\label{20}
\frac{3}{2}{\square ^{(4)}}\Psi - \frac{3}{2}\Psi'' - \frac{3}{2}\frac{{a'}}{a}\Psi'
+ \kappa _5^2\phi {'^2}{\mathcal{L}_{XX}}{a^{ - 2}}\phi {'^2}\Psi \nonumber\\
&&= 2\kappa _5^2{\mathcal{L}_{X}}\phi '\Phi ' - \kappa _5^2\phi {'^2}{\mathcal{L}_{XX}}{a^{ - 2}}\phi '\Phi ' + \kappa _5^2\phi {'^2}{\mathcal{L}_{X\phi }}\Phi.
\end{eqnarray}

Note that the use of gauge-invariant variables here is equivalent to taking the so-called longitudinal gauge gauge, which takes $\psi=0$, $B=0$ and $C_\mu=0$. Under this gauge, $\{\Xi,~\Psi,~\Phi,~v_\mu\}$ simply reduce to $\{h_{rr},~A,~\phi,~G_\mu\}$, respectively. So, the final equations for $\{h_{rr},~A,~\phi\}$ are nothing but
Eqs.~\eqref{18}-\eqref{20}. What we need to do is simply replace $\{\Xi,~\Psi,~\Phi\}$ to $\{h_{rr},~A,~\phi\}$.
Both methods completely eliminate the gauge freedoms. In fact, for any gauge that completely eliminates the gauge freedoms, we can always construct the corresponding gauge-invariant variables. A good choice of gauge usually helps us to simplify the perturbation equations.

Eliminating $\Xi$, $\Phi$ and $\mathcal{L}_{X\phi}$ by using Eqs.~\eqref{18} and \eqref{19} and background equations, correspondingly, one can reexpress Eq.~\eqref{20} as
\begin{eqnarray}
  &&{\square ^{(4)}}\Psi +(1+2f)\Psi''\nonumber\\
  &&+(1+2f)\left[\partial _y\ln  \left(\frac{a^3}{\mathcal{L}_X(\phi ')^2}\right)\right]\Psi '\nonumber\\
  &&+2(1+2f)\mathcal{H} \left[\partial _y\ln  \left(\frac{\mathcal{H}^2}{\mathcal{L}_X(\phi ')^2}\right)\right]\Psi=0,
\end{eqnarray}
where $\mathcal{H}\equiv a'/a$.
For non-negative $\mathcal{L}_X$, the above equation takes a more compact form:
\begin{eqnarray}
\label{schro}
{\square ^{(4)}}\Psi+\gamma\Psi''-\gamma z\left(z^{-1}\right)''\Psi =0,
\end{eqnarray} after redefining $\Psi \to a^{-3/2}\mathcal{L}_X^{1/2}\phi '\Psi$. Here
\begin{equation}
\label{zf}
z=a^{3/2}\frac{\phi '}{\mathcal{H} }\mathcal{L}_X^{1/2},\quad
\gamma=1+2\frac{\mathcal{L}_{XX} X}{\mathcal{L}_X}.
\end{equation}
If further $\gamma>0$, then we can use the Regge-Wheeler ``tortoise'' coordinate $r^{\ast}$, such that
\begin{equation}
\label{RWcoord}
\frac{dr^{\ast}}{dr}\equiv\gamma^{-1/2},
\end{equation}
to rewrite Eq.~\eqref{schro} as
\begin{eqnarray}
&&{\square ^{(4)}}\Psi+\ddot{\Psi}-\frac{\dot{\gamma}}{2\gamma}\dot{\Psi}\nonumber\\
&-& z\left(z^{-1}\right)^{\centerdot\centerdot}\Psi
+\frac{\dot{\gamma}}{2\gamma} \left(z^{-1}\right)^{\centerdot}z\Psi=0.
\end{eqnarray}
Here, we have used dots to denote the derivative with respect to $r^{\ast}$, for example, $\dot{\Psi}\equiv\frac{d \Psi}{dr^{\ast}}$.
After a further redefinition of the field $\Psi \to \gamma^{1/4}\Psi$, we finally obtain
\begin{eqnarray}
\label{eqPhi}
{\square ^{(4)}}\Psi+\ddot{\Psi}-\Psi\theta \left(\theta ^{-1}\right)^{\centerdot\centerdot}=0,
\end{eqnarray}
where $\theta \equiv\gamma^{1/4}z $.
The massive modes of $\Psi$ satisfy the following equation:
\begin{equation}
\label{Psi}
\mathcal{A}^\dagger\mathcal{A}\Psi_m=m_\Psi^2\Psi_m,
\end{equation}
with
\begin{equation}
\label{Adagger}
\mathcal{A}=\frac{d}{dr^{\ast}}+\frac{\dot{\theta}}{\theta},\quad
\mathcal{A}^\dagger=-\frac{d}{dr^{\ast}}+\frac{\dot{\theta}}{\theta}.
\end{equation}
Obviously, the zero mode takes the form $\Psi_0\propto \theta^{-1}$.

So far, we have shown that after a series of redefinitions of both the field $\Psi$ and the coordinate, we have transformed the master equation of $\Psi$ into the Schr\"odinger-like equation \eqref{eqPhi}. The factorization of the Schr\"odinger-like equation ensures that $m_\Psi^2\geq0$. Consequently, we can say that any solution with $\mathcal{L}_X>0$ and $\gamma>0$ is stable. However, from the point of view of quadratic action, $\Psi$ and many other gauge-invariant quantities, despite satisfing some simple equations, are not the canonical normal modes that diagonalize the quadratic action~\cite{Giovannini2003}. As we will see in the next section, only by considering the quadratic action can we obtain the normal modes of the perturbations. It is the normal modes that should be considered as the dynamical variables in the quantization procedure.

\section{Quadratic action and the normal modes}
\label{sec4}
Following a similar procedure to Ref.~\cite{Giovannini2003}, we obtain the second-order expansion of the gravitational Lagrangian:
\begin{eqnarray}
\label{Lg}
&&{\delta ^{(2)}}{\mathcal{L}_{\textrm{gravity}}}= \frac{1}{2}\sqrt { - g} {a^{ - 2}}\bigg\{
 {\partial _M}{h_{NP}}{\partial ^P}{h^{MN}}
 \nonumber \\
   && - {\partial ^M}h{\partial ^N}{h_{MN}}
  - \frac{1}{2}{\partial _P}{h_{MN}}{\partial ^P}{h^{MN}}
   + \frac{1}{2}{\partial ^M}h{\partial _M}h \nonumber \\
   &&
   + 3\frac{{a'}}{a}h{\partial ^M}{h_{Mr}}+ {\left( {\frac{{a'}}{a}} \right)^2}12{h^{Mr}}{h_{Mr}} \nonumber \\
   &&
   + \frac{{a''}}{a}\Big(6{h_{rr}}h
    - \frac{3}{2}{h^2} + 3{h^{MN}}{h_{MN}} - 6{h^{Mr}}{h_{Mr}}\Big) \bigg\}.
\end{eqnarray}

Likewise, the second-order perturbation of the matter Lagrangian density ${\mathcal{L}_{\textrm{matter}}} = \sqrt { - g} \mathcal{L}(\phi ,X)$ is given by
\begin{eqnarray}
\label{Lm}
&&{\delta ^{(2)}}{\mathcal{L}_\textrm{matter}}= \frac{1}{2}{a^3}\bigg\{ \nonumber \\
    & & \frac{1}{4}{a^2}\mathcal{L}{h^2} - \frac{1}{2}{a^2}\mathcal{L}{h_{MN}}{h^{MN}} + {a^2}{\mathcal{L}_{\phi \phi }}{(\delta \phi )^2}\nonumber \\
    & &+ {a^{ - 2}}{\mathcal{L}_{XX}}{(\phi ')^2}\Big(\delta \phi '-\frac12 \phi '{h_{rr}}\Big)^2\nonumber \\
   & &+ {\mathcal{L}_X}\Big[\frac{1}{2}{(\phi ')^2}h{h_{rr}}
    + 2{h^{Mr}}\phi '{\partial _M}\delta \phi
    + \phi 'h'\delta \phi  \nonumber \\
   & &
    - {\partial ^M}\delta \phi {\partial _M}\delta \phi
    - {(\phi ')^2}{h^{Mr}}{h_{Mr}}\Big]\nonumber \\
    & &
  -2{\mathcal{L}_{\phi X}}\phi '\delta \phi\left(
    \delta \phi '-\frac12{\phi '}{h_{rr}}
      \right)\bigg\}.
\end{eqnarray}
We have eliminated ${\mathcal{L}_\phi }$ by using the background equation \eqref{EOM}.
The full quadratic Lagrangian density can be obtained by combining Eqs.~\eqref{Lg} and \eqref{Lm}:
\begin{eqnarray}
 &&  {\delta ^{(2)}}{\mathcal{L}_{\textrm{total}}}=
  \frac{1}{2}{a^{3}}\bigg\{
   {\partial _M}{h_{NP}}{\partial ^P}{h^{MN}}\nonumber \\
   &&
    - {\partial ^M}h{\partial ^N}{h_{MN}}
     - \frac{1}{2}{\partial _P}{h_{MN}}{\partial ^P}{h^{MN}}
     + \frac{1}{2}{\partial ^P}h{\partial _P}h \nonumber \\
   &&
   + 3\frac{{a'}}{a}h{\partial ^\mu }{h_{\mu r}} - 3\frac{{a'}}{a}{h_{rr}}h' + 2\kappa _5^2{a^2}{\mathcal{L}_{\phi \phi }}{(\delta \phi )^2} \nonumber \\
   &&
   +2\kappa _5^2{a^{ - 2}}{\mathcal{L}_{XX}}{(\phi ')^2}\Big(\delta \phi '-\frac12 \phi '{h_{rr}}\Big)^2\nonumber \\
   &&
   + 2\kappa _5^2{\mathcal{L}_X}\Big[ 2{h^{Mr}}\phi '{\partial _M}\delta \phi  + \phi 'h'\delta \phi  - {\partial ^M}\delta \phi {\partial _M}\delta \phi \Big] \nonumber \\
   &&
   -4\kappa _5^2{\mathcal{L}_{\phi X}}\phi '\delta \phi\left(
    \delta \phi '-\frac12{\phi '}{h_{rr}}
      \right) \bigg\}.
\end{eqnarray}

Plugging Eq.~\eqref{decomposition} into the above equation, ${\delta ^{(2)}}{\mathcal{L}_{\textrm{total}}}$ decouples into several building blocks, for example, the vector and tensor sections are
\begin{eqnarray}
\label{37}
  {\delta ^{(2)}}{\mathcal{L}_{\textrm{vector}}}& =& \frac{1}{2} {\hat{v}^\mu }{\square ^{(4)}}{\hat{v}_\mu },   \\
  {\delta ^{(2)}}{\mathcal{L}_{\textrm{tensor}}}
  &= & \frac{1}{4}\hat{D}^{\mu \nu }
  \bigg\{\square ^{(4)}{\hat{D}_{\mu \nu }} + \hat{D}_{\mu \nu }'' - \frac{{({a^{\frac{3}{2}}})''}}{{{a^{\frac{3}{2}}}}}{\hat{D}^{\mu \nu }}\bigg\},
\end{eqnarray}
with
\begin{equation}
\hat{v}^\mu={a^{\frac{3}{2}}}{v^\mu},\quad {\hat{D}^{\mu \nu }} = {a^{\frac{3}{2}}}{D^{\mu \nu }}.
\end{equation}

For the tensor section, the normal mode satisfies the same equation we obtained in Sec. \ref{sec3}, so we will not repeat the discussions here.

It is worth noting that for the vector section, we can recover only Eq.~\eqref{vc1}. Some authors argued that Eq.~\eqref{vc2} might come from the following Lagrangian~\cite{Giovannini2001a}:
\begin{eqnarray}
\label{vecconje}
a^3{\partial^{(\mu }}C^{\nu )}  \Big\{3\frac{{a'}}{a}{\partial _{(\mu }}{G_{\nu )}} + {\partial _{(\mu }}G{'_{\nu )}} \Big\}.
\end{eqnarray}
However, as shown in Eq.~\eqref{37}, we did not obtain such a Lagrangian. So for the vector section, there is a discrepancy between the equations obtained by using different methods. A similar discrepancy was also found in cosmology in both Einstein gravity and Ho\v{r}ava-Lifshitz gravity~\cite{GongKohSasaki2010}.

Therefore, from the point of view of action, we also conclude that only the zero mode of the vector perturbation survives, so our model is stable against vector perturbations. However, from the quadratic action of vector perturbation, we cannot tell whether the vector zero mode is localizable. The reason for this discrepancy is still not clear to us, so we would like to make further investigation on this issue. But in this paper, our main interest lies in the scalar section.

The first building block of the scalar perturbation relates to $\psi=F-\frac12B'$:
\begin{eqnarray}
&&{\delta ^{(2)}}{\mathcal{L}_{\textrm{scalar-1}}}\nonumber \\&& = {a^3}
 \Big\{ 3\frac{{a'}}{a}{h_{rr}} - 3A' - 2\kappa _5^2{\mathcal{L}_X}\phi '\delta \phi \Big\} {\square ^{(4)}}\psi,
\end{eqnarray}
which leads to the following constraint
\begin{eqnarray}
\label{hrr}
3\frac{{a'}}{a}{h_{rr}} - 3A' - 2\kappa _5^2{\mathcal{L}_X}\phi '\delta \phi=0.
\end{eqnarray}
This is nothing but Eq.~\eqref{19}.
Another part of the scalar Lagrangian contains $A$, $h_{rr}$, and $\delta\phi$:
\begin{eqnarray}
 & & {\delta ^{(2)}}{\mathcal{L}_{\textrm{scalar-2}}}\nonumber \\&&
 =\frac{1}{2}{a^{ 3}} \Big\{
 - 3A{\square ^{(4)}}A
 - 3{h_{rr}}{\square ^{(4)}}A
 +2\kappa _5^2{\mathcal{L}_X}\delta \phi \square ^{(4)}\delta \phi \nonumber \\&&
  + 6A'A'
   - 3\frac{{a'}}{a}{h_{rr}}(h_{rr}' + 4A')
   + 2\kappa _5^2{a^2}{\mathcal{L}_{\phi \phi }}{(\delta \phi )^2}\nonumber \\&&
   + 4\kappa _5^2{\mathcal{L}_X}{h_{rr}}\phi '\delta \phi'
    + 2\kappa _5^2{\mathcal{L}_X}\phi '(h_{rr}' + 4A')\delta \phi
    \nonumber \\&&
   - 2\kappa _5^2{\mathcal{L}_X}(\delta \phi' )^2
   -4\kappa _5^2{\mathcal{L}_{\phi X}}\phi '\delta \phi\left(
    \delta \phi '-\frac12{\phi '}{h_{rr}}
      \right) \nonumber \\&&
   + 2\kappa _5^2{a^{ - 2}}{\mathcal{L}_{XX}}{(\phi ')^2}\Big(\delta \phi '-\frac12 \phi '{h_{rr}}\Big)^2 \Big\}.
\end{eqnarray}
To diagonalize ${\delta ^{(2)}}{\mathcal{L}_{\textrm{scalar-2}}}$, we firstly eliminate $h_{rr}$ by using Eq.~\eqref{hrr}. Then, we introduce a gauge-invariant variable which combines $A$ and $\delta\phi$:
\begin{eqnarray}
\label{G}
\mathcal{G} &\equiv& a^{3/2}\sqrt{\mathcal{L}_X}
 \Big(2\delta \phi  - \frac{{\phi '}}{\mathcal{H}}A \Big).
\end{eqnarray}
Here ${\mathcal{L}_X}>0$ is required.
Naively, after the elimination of $\delta \phi$, $ {\delta ^{(2)}}{\mathcal{L}_{\textrm{scalar-2}}}$ should be expressed in terms of $A$ and $\mathcal{G}$. However, since we did not take any gauge, $ {\delta ^{(2)}}{\mathcal{L}_{\textrm{scalar-2}}}$ is still gauge invariant. That means all terms that contain $A$ and its derivatives must be vanished, because these terms cannot be gauge invariant (one can show that after a long but straightforward calculation, all terms with $A$ are indeed vanished). Thus, we get a Lagrangian density with only $\mathcal{G}$, namely,
\begin{eqnarray}
{\delta ^{(2)}}{S_\mathcal{G}} &= &\frac{1}{4}\kappa _5^2\int d^4xdr\nonumber\\
 &\times& \left\{
   \mathcal{G}{\square ^{(4)}}\mathcal{G} +V(r)\mathcal{G}^2
+(1+2f) \mathcal{G}\mathcal{G}''\right\},
\end{eqnarray}
with
\begin{equation}
V(r)=-\frac{z''}{z}-\left(\frac{z''}{z}+\frac{(zf)''}{zf}\right)f.
\end{equation}
Here $z$ is defined in Eq.~\eqref{zf}, and
\begin{equation}
f=\frac{\mathcal{L}_{XX} X}{\mathcal{L}_X}.
\end{equation}

If $\gamma=1+2f>0$, we can use the coordinate $r^{\ast}$ defined in Eq.~\eqref{RWcoord} to rewrite the quadratic action as
\begin{eqnarray}
{\delta ^{(2)}}{S_\mathcal{G}} &= &\frac{1}{4}\kappa _5^2\int d^4xdr^{\ast}{\sqrt{1+2f}}\nonumber\\
 &\times &
 \left\{
   \mathcal{G}{\square ^{(4)}}\mathcal{G} +U(r^{\ast})\mathcal{G}^2
+ \mathcal{G} \ddot{\mathcal{G}}\right\},
\end{eqnarray}
with
\begin{equation}
U(r^{\ast})\equiv V(r^{\ast})+\frac12\frac{\ddot{f}}{1+2f}-\left(\frac{\dot{f}}{1+2f}\right)^2.
\end{equation}
Thus, the scalar normal mode should be
\begin{equation}
\hat{\mathcal{G}}=\frac{\kappa_5}{2}(1+2f)^{1/4}\mathcal{G},
\end{equation}
and the corresponding action is
\begin{equation}
 {\delta ^{(2)}}{S_{\hat{\mathcal{G}}}} = \int d^4xdr^{\ast}\hat{\mathcal{G}}
 \left\{
   {\square ^{(4)}}\hat{\mathcal{G}}+ \ddot{\hat{\mathcal{G}}} -\frac{\ddot{\theta}}{\theta}\hat{\mathcal{G}}
\right\}.
\end{equation}
Again, we defined $\theta \equiv\gamma^{1/4}z $.
Obviously, for the standard case $\gamma=1$ and $r^{\ast}=r$, our result reduces to the one given in Ref.~\cite{Giovannini2003}.

Note that one can rewrite $\mathcal{G}$ in Eq.~\eqref{G} in terms of gauge-invariant variables:
\begin{eqnarray}
\mathcal{G} =a^{3/2}\sqrt{\mathcal{L}_X} \Big(2 \Phi  - \frac{{\phi '}}{\mathcal{H}}\Psi \Big).
\end{eqnarray}
We see that $\mathcal{G}$ explicitly contains the matter perturbation $\Phi$. In contrast, neither $\Psi$ nor $\Phi$ can diagonalize the quadratic action. So, it is $\mathcal{G}$ rather than $\Psi$ that should be served as the normal mode in the quantization. This problem was pointed out previously~\cite{Giovannini2003}. On the other hand, one can also derive the same equation for $\hat{\mathcal{G}}$, namely, Eq.~\eqref{hatG}, by simply combining the equations of $\Psi$ and $\Phi$ (see Ref.~\cite{Giovannini2001a} for details).

\section{Stability and localization of the canonical scalar zero mode}
\label{sec5}
From the bilinear action of $\hat{\mathcal{G}}$, we know that for $\mathcal{L}_X>0$ and $1+2f>0$, the scalar normal mode satisfies a Schr\"odinger-like equation
\begin{equation}
\label{schroScalar}
-\ddot{\hat{\mathcal{G}}} +\frac{\ddot{\theta}}{\theta}\hat{\mathcal{G}}
={\square ^{(4)}}\hat{\mathcal{G}}.
\end{equation}
The massive modes of $\hat{\mathcal{G}}$ are described by the following equation
\begin{equation}
\label{hatG}
\mathcal{A}\mathcal{A}^\dagger\hat{\mathcal{G}}_m=m_{\hat{\mathcal{G}}}^2\hat{\mathcal{G}}_m,
\end{equation} where $\mathcal{A}$ and $\mathcal{A}^\dagger$ are defined in Eq.~\eqref{Adagger}. According to the supersymmetric quantum mechanics, we know that $m_{\hat{\mathcal{G}}}^2$ is non-negative, and consequently, any solution of models with $\mathcal{L}_X>0$ and $1+2f>0$ should be stable under the scalar perturbation. Moreover, according to Eqs.~\eqref{Psi} and \eqref{hatG}, we find that $\hat{\mathcal{G}}_m$ and $\Psi_m$ are superpartners, that means their mass spectra are related~\cite{Giovannini2001a}.

The normalization condition for the canonical zero mode $\hat{\mathcal{G}}_0=K\theta(r^{\ast})$ is
\begin{eqnarray}
&&K^2\int dr^{\ast} |\theta|^2=K^2\int dr^{\ast} (1+2f)^{1/2}(z(r^{\ast}))^2\nonumber\\
&&=K^2\int dr\mathcal{L}_X a^{3}\frac{\phi '^2}{\mathcal{H}^2 }\nonumber\\
&&=3\frac{K^2}{\kappa_5^2}\int dr a^{3}\frac{{\mathcal{H}^2} - \mathcal{H}'}{\mathcal{H}^2 }=1.
\end{eqnarray}
Comparing this condition to the one obtained in the standard case $\mathcal{L}=X-V$~\cite{Giovannini2001a}, we conclude that the modification of the scalar Lagrangian does not affect the localization of the scalar zero mode. This result does make sense, because as we can see in the standard case, the localization of the scalar zero mode is determined only by the warp factor (or the geometry of space-time) rather than the matter field~\cite{Giovannini2001a}. Thus, as stated in Ref.~\cite{Giovannini2003}, the scalar zero mode is not localizable, because it is divergent either at $r\to 0$ or at $r\to\infty$.

The massive scalar spectrum would be different as compared to
the standard case, however. Thus, one expects to see different modification to the four-dimensional gravity. We will leave this problem to our future work.

\section{Conclusions}
\label{sec6}
In this paper, we investigated the linear perturbation of a class of thick $K$-branes. We derived the master equations for linear perturbations from the point of view of both dynamical equations and quadratic action. The canonical normal mode of the scalar perturbation turns out to be a combination of both the metric and field perturbations. The localization of the scalar zero mode depends only on the geometry of the space-time, rather than the explicit form of the Lagrangian of the scalar field. Therefore, the scalar perturbation is nonlocalizable in general. The massive modes of the canonical scalar perturbation satisfy a Schr\"odinger-like equation, provided $\mathcal{L}_X>0$ and $\gamma=1+2X\frac{\mathcal{L}_{XX} }{\mathcal{L}_X}>0$. Note that the first condition $\mathcal{L}_X>0$ is required as the consequence of the null energy condition $T_{MN}n^M n^N\geq0$, for an arbitrary null vector $n^M $ such that $n^M n_M=0$. The decomposition of the Schr\"odinger-like equation indicates that there is no tachyon in the mass spectrum. So thick $K$-brane solutions with $\mathcal{L}_X>0$ and $\gamma>0$ are generally stable under linear scalar perturbation.

Besides, the modification of the matter Lagrangian does not affect the tensor and vector sections. So, after linearizing the Einstein equations, we get the same conclusions as the standard case, namely, the vector perturbation has only zero mode, while the tensor perturbation has both zero and massive modes. In order to have a finite four-dimensional Planck constant, the localization of the tensor zero mode must be required, as a consequence, the vector perturbation cannot be localized due to Eq.~\eqref{vc2}. However, from the point of view of quadratic action, Eq.~\eqref{vc2} is absent due to some unknown reasons. A further investigation on this issue is necessary, but we would like to leave it to the future work.

\section{Acknowledgement}
We would like to thank Evslin Jarah, Huan Hsin Tseng, and Yu-Hsiang Lin for helpful discussions on the gravitational perturbations and other issues. Especially, we thank the referee for crucial comments and suggestions, which helped us to improve the original manuscript. This work was supported by the Program for New Century Excellent Talents in University, the National Natural Science Foundation of China (Grant No. 11075065), and the Huo Ying-Dong Education Foundation of Chinese Ministry of Education (Grant No. 121106), and the Fundamental Research Funds for the Central Universities (Grant No. lzujbky-2013-18). Y. Z. was supported by the Scholarship Award for Excellent Doctoral Student granted by Ministry of Education.

%


\begin{thebibliography}{38}%
\makeatletter
\providecommand \@ifxundefined [1]{%
 \@ifx{#1\undefined}
}%
\providecommand \@ifnum [1]{%
 \ifnum #1\expandafter \@firstoftwo
 \else \expandafter \@secondoftwo
 \fi
}%
\providecommand \@ifx [1]{%
 \ifx #1\expandafter \@firstoftwo
 \else \expandafter \@secondoftwo
 \fi
}%
\providecommand \natexlab [1]{#1}%
\providecommand \enquote  [1]{``#1''}%
\providecommand \bibnamefont  [1]{#1}%
\providecommand \bibfnamefont [1]{#1}%
\providecommand \citenamefont [1]{#1}%
\providecommand \href@noop [0]{\@secondoftwo}%
\providecommand \href [0]{\begingroup \@sanitize@url \@href}%
\providecommand \@href[1]{\@@startlink{#1}\@@href}%
\providecommand \@@href[1]{\endgroup#1\@@endlink}%
\providecommand \@sanitize@url [0]{\catcode `\\12\catcode `\$12\catcode
  `\&12\catcode `\#12\catcode `\^12\catcode `\_12\catcode `\%12\relax}%
\providecommand \@@startlink[1]{}%
\providecommand \@@endlink[0]{}%
\providecommand \url  [0]{\begingroup\@sanitize@url \@url }%
\providecommand \@url [1]{\endgroup\@href {#1}{\urlprefix }}%
\providecommand \urlprefix  [0]{URL }%
\providecommand \Eprint [0]{\href }%
\@ifxundefined \urlstyle {%
  \providecommand \doi  [0]{\begingroup \@sanitize@url \@doi}%
  \providecommand \@doi [1]{\endgroup \@@startlink {\doibase
  #1}doi:\discretionary {}{}{}#1\@@endlink }%
}{%
  \providecommand \doi  [0]{doi:\discretionary{}{}{}\begingroup
  \urlstyle{rm}\Url }%
}%
\providecommand \doibase [0]{http://dx.doi.org/}%
\providecommand \Doi [0]{\begingroup \@sanitize@url \@Doi }%
\providecommand \@Doi  [1]{\endgroup\@@startlink{\doibase#1}\@@Doi}%
\providecommand \@@Doi [1]{#1\@@endlink}%
\providecommand \selectlanguage [0]{\@gobble}%
\providecommand \bibinfo  [0]{\@secondoftwo}%
\providecommand \bibfield  [0]{\@secondoftwo}%
\providecommand \translation [1]{[#1]}%
\providecommand \BibitemOpen [0]{}%
\providecommand \bibitemStop [0]{}%
\providecommand \bibitemNoStop [0]{.\EOS\space}%
\providecommand \EOS [0]{\spacefactor3000\relax}%
\providecommand \BibitemShut  [1]{\csname bibitem#1\endcsname}%
\bibitem [{\citenamefont {Armendariz-Picon}\ \emph {et~al.}(1999)\citenamefont
  {Armendariz-Picon}, \citenamefont {Damour},\ and\ \citenamefont
  {Mukhanov}}]{Armendariz-PiconDamourMukhanov1999}%
  \BibitemOpen
  \bibfield  {author} {\bibinfo {author} {\bibfnamefont {C.}~\bibnamefont
  {Armendariz-Picon}}, \bibinfo {author} {\bibfnamefont {T.}~\bibnamefont
  {Damour}}, \ and\ \bibinfo {author} {\bibfnamefont {V.~F.}\ \bibnamefont
  {Mukhanov}},\ }\Doi {10.1016/S0370-2693(99)00603-6} {\bibfield  {journal}
  {\bibinfo  {journal} {Phys. Lett. B}\ }\textbf {\bibinfo {volume} {458}},\
  \bibinfo {pages} {209} (\bibinfo {year} {1999})} \BibitemShut
  {NoStop}%
\bibitem [{\citenamefont {Garriga}\ and\ \citenamefont
  {Mukhanov}(1999)}]{GarrigaMukhanov1999}%
  \BibitemOpen
  \bibfield  {author} {\bibinfo {author} {\bibfnamefont {J.}~\bibnamefont
  {Garriga}}\ and\ \bibinfo {author} {\bibfnamefont {V.~F.}\ \bibnamefont
  {Mukhanov}},\ }\Doi {10.1016/S0370-2693(99)00602-4} {\bibfield  {journal}
  {\bibinfo  {journal} {Phys. Lett. B}\ }\textbf {\bibinfo {volume} {458}},\
  \bibinfo {pages} {219} (\bibinfo {year} {1999})}
  \BibitemShut {NoStop}%
\bibitem [{\citenamefont {Armendariz-Picon}\ \emph {et~al.}(2001)\citenamefont
  {Armendariz-Picon}, \citenamefont {Mukhanov},\ and\ \citenamefont
  {Steinhardt}}]{Armendariz-PiconMukhanovSteinhardt2001}%
  \BibitemOpen
  \bibfield  {author} {\bibinfo {author} {\bibfnamefont {C.}~\bibnamefont
  {Armendariz-Picon}}, \bibinfo {author} {\bibfnamefont {V.~F.}\ \bibnamefont
  {Mukhanov}}, \ and\ \bibinfo {author} {\bibfnamefont {P.~J.}\ \bibnamefont
  {Steinhardt}},\ }\Doi {10.1103/PhysRevD.63.103510} {\bibfield  {journal}
  {\bibinfo  {journal} {Phys. Rev. D}\ }\textbf {\bibinfo {volume} {63}},\
  \bibinfo {pages} {103510} (\bibinfo {year} {2001})}
  \BibitemShut {NoStop}%
\bibitem [{\citenamefont {Rubakov}\ and\ \citenamefont
  {Shaposhnikov}(1983)}]{RubakovShaposhnikov1983}%
  \BibitemOpen
  \bibfield  {author} {\bibinfo {author} {\bibfnamefont {V.~A.}\ \bibnamefont
  {Rubakov}}\ and\ \bibinfo {author} {\bibfnamefont {M.~E.}\ \bibnamefont
  {Shaposhnikov}},\ }\href@noop {} {\bibfield  {journal} {\bibinfo  {journal}
  {Phys. Lett. B}\ }\textbf {\bibinfo {volume} {125}},\ \bibinfo {pages} {136}
  (\bibinfo {year} {1983})}\BibitemShut {NoStop}%
\bibitem [{\citenamefont {Randall}\ and\ \citenamefont
  {Sundrum}(1999){\natexlab{a}}}]{RandallSundrum1999}%
  \BibitemOpen
  \bibfield  {author} {\bibinfo {author} {\bibfnamefont {L.}~\bibnamefont
  {Randall}}\ and\ \bibinfo {author} {\bibfnamefont {R.}~\bibnamefont
  {Sundrum}},\ }\Doi {10.1103/PhysRevLett.83.3370} {\bibfield  {journal}
  {\bibinfo  {journal} {Phys. Rev. Lett.}\ }\textbf {\bibinfo {volume} {83}},\
  \bibinfo {pages} {3370} (\bibinfo {year} {1999}{\natexlab{a}})} \BibitemShut
  {NoStop}%
\bibitem [{\citenamefont {Randall}\ and\ \citenamefont
  {Sundrum}(1999){\natexlab{b}}}]{RandallSundrum1999a}%
  \BibitemOpen
  \bibfield  {author} {\bibinfo {author} {\bibfnamefont {L.}~\bibnamefont
  {Randall}}\ and\ \bibinfo {author} {\bibfnamefont {R.}~\bibnamefont
  {Sundrum}},\ }\Doi {10.1103/PhysRevLett.83.4690} {\bibfield  {journal}
  {\bibinfo  {journal} {Phys. Rev. Lett.}\ }\textbf {\bibinfo {volume} {83}},\
  \bibinfo {pages} {4690} (\bibinfo {year} {1999}{\natexlab{b}})} \BibitemShut
  {NoStop}%
\bibitem [{\citenamefont {DeWolfe}\ \emph {et~al.}(2000)\citenamefont
  {DeWolfe}, \citenamefont {Freedman}, \citenamefont {Gubser},\ and\
  \citenamefont {Karch}}]{DeWolfeFreedmanGubserKarch2000}%
  \BibitemOpen
  \bibfield  {author} {\bibinfo {author} {\bibfnamefont {O.}~\bibnamefont
  {DeWolfe}}, \bibinfo {author} {\bibfnamefont {D.~Z.}\ \bibnamefont
  {Freedman}}, \bibinfo {author} {\bibfnamefont {S.~S.}\ \bibnamefont
  {Gubser}}, \ and\ \bibinfo {author} {\bibfnamefont {A.}~\bibnamefont
  {Karch}},\ }\Doi {10.1103/PhysRevD.62.046008} {\bibfield  {journal} {\bibinfo
   {journal} {Phys. Rev. D}\ }\textbf {\bibinfo {volume} {62}},\ \bibinfo
  {pages} {046008} (\bibinfo {year} {2000})} \BibitemShut
  {NoStop}%
\bibitem [{\citenamefont {Gremm}(2000){\natexlab{a}}}]{Gremm2000}%
  \BibitemOpen
  \bibfield  {author} {\bibinfo {author} {\bibfnamefont {M.}~\bibnamefont
  {Gremm}},\ }\Doi {10.1016/S0370-2693(00)00303-8} {\bibfield  {journal}
  {\bibinfo  {journal} {Phys. Lett. B}\ }\textbf {\bibinfo {volume} {478}},\
  \bibinfo {pages} {434} (\bibinfo {year} {2000}{\natexlab{a}})} \BibitemShut
  {NoStop}%
\bibitem [{\citenamefont {Gremm}(2000){\natexlab{b}}}]{Gremm2000a}%
  \BibitemOpen
  \bibfield  {author} {\bibinfo {author} {\bibfnamefont {M.}~\bibnamefont
  {Gremm}},\ }\Doi {10.1103/PhysRevD.62.044017} {\bibfield  {journal} {\bibinfo
   {journal} {Phys. Rev. D}\ }\textbf {\bibinfo {volume} {62}},\ \bibinfo
  {pages} {044017} (\bibinfo {year} {2000}{\natexlab{b}})} \BibitemShut
  {NoStop}%
\bibitem [{\citenamefont {Csaki}\ \emph {et~al.}(2000)\citenamefont {Csaki},
  \citenamefont {Erlich}, \citenamefont {Hollowood},\ and\ \citenamefont
  {Shirman}}]{CsakiErlichHollowoodShirman2000}%
  \BibitemOpen
  \bibfield  {author} {\bibinfo {author} {\bibfnamefont {C.}~\bibnamefont
  {Csaki}}, \bibinfo {author} {\bibfnamefont {J.}~\bibnamefont {Erlich}},
  \bibinfo {author} {\bibfnamefont {T.~J.}\ \bibnamefont {Hollowood}}, \ and\
  \bibinfo {author} {\bibfnamefont {Y.}~\bibnamefont {Shirman}},\ }\Doi
  {10.1016/S0550-3213(00)00271-6} {\bibfield  {journal} {\bibinfo  {journal}
  {Nucl. Phys. B}\ }\textbf {\bibinfo {volume} {581}},\ \bibinfo {pages} {309}
  (\bibinfo {year} {2000})} \BibitemShut {NoStop}%
\bibitem [{\citenamefont {Giovannini}(2001)}]{Giovannini2001a}%
  \BibitemOpen
  \bibfield  {author} {\bibinfo {author} {\bibfnamefont {M.}~\bibnamefont
  {Giovannini}},\ }\Doi {10.1103/PhysRevD.64.064023} {\bibfield  {journal}
  {\bibinfo  {journal} {Phys. Rev. D}\ }\textbf {\bibinfo {volume} {64}},\
  \bibinfo {pages} {064023} (\bibinfo {year} {2001})}
  \BibitemShut {NoStop}%
\bibitem [{\citenamefont {Giovannini}(2003)}]{Giovannini2003}%
  \BibitemOpen
  \bibfield  {author} {\bibinfo {author} {\bibfnamefont {M.}~\bibnamefont
  {Giovannini}},\ }\Doi {10.1088/0264-9381/20/6/303} {\bibfield  {journal}
  {\bibinfo  {journal} {Classical Quantum Gravity} \ }\textbf {\bibinfo
  {volume} {20}},\ \bibinfo {pages} {1063} (\bibinfo {year} {2003})}
  \BibitemShut {NoStop}%
\bibitem [{\citenamefont {Giovannini}(2002)}]{Giovannini2002}%
  \BibitemOpen
  \bibfield  {author} {\bibinfo {author} {\bibfnamefont {M.}~\bibnamefont
  {Giovannini}},\ }\Doi {10.1103/PhysRevD.65.064008} {\bibfield  {journal}
  {\bibinfo  {journal} {Phys. Rev. D}\ }\textbf {\bibinfo {volume} {65}},\
  \bibinfo {pages} {064008} (\bibinfo {year} {2002})}
  \BibitemShut {NoStop}%
\bibitem [{\citenamefont {Rubakov}(2001)}]{Rubakov2001}%
  \BibitemOpen
  \bibfield  {author} {\bibinfo {author} {\bibfnamefont {V.~A.}\ \bibnamefont
  {Rubakov}},\ }\Doi {10.1070/PU2001v044n09ABEH001000} {\bibfield  {journal}
  {\bibinfo  {journal} {Phys. Usp.}\ }\textbf {\bibinfo {volume} {44}},\
  \bibinfo {pages} {871} (\bibinfo {year} {2001})} \BibitemShut
  {NoStop}%
\bibitem [{\citenamefont {Csaki}(2004)}]{Csaki2004}%
  \BibitemOpen
  \bibfield  {author} {\bibinfo {author} {\bibfnamefont {C.}~\bibnamefont
  {Csaki}},\ }\href@noop {} {\enquote {\bibinfo {title} {Tasi lectures on extra
  dimensions and branes},}\ } (\bibinfo {year} {2004}),\ \Eprint
  {http://arxiv.org/abs/hep-ph/0404096} {arXiv:hep-ph/0404096} \BibitemShut
  {NoStop}%
\bibitem [{\citenamefont {Shifman}(2010)}]{Shifman2010}%
  \BibitemOpen
  \bibfield  {author} {\bibinfo {author} {\bibfnamefont {M.}~\bibnamefont
  {Shifman}},\ }\Doi {10.1142/S0217751X10048548} {\bibfield  {journal}
  {\bibinfo  {journal} {Int. J. Mod. Phys. A}\ }\textbf {\bibinfo {volume}
  {25}},\ \bibinfo {pages} {199} (\bibinfo {year} {2010})}\BibitemShut
  {NoStop}%
\bibitem [{\citenamefont {Dzhunushaliev}\ \emph {et~al.}(2010)\citenamefont
  {Dzhunushaliev}, \citenamefont {Folomeev},\ and\ \citenamefont
  {Minamitsuji}}]{DzhunushalievFolomeevMinamitsuji2010}%
  \BibitemOpen
  \bibfield  {author} {\bibinfo {author} {\bibfnamefont {V.}~\bibnamefont
  {Dzhunushaliev}}, \bibinfo {author} {\bibfnamefont {V.}~\bibnamefont
  {Folomeev}}, \ and\ \bibinfo {author} {\bibfnamefont {M.}~\bibnamefont
  {Minamitsuji}},\ }\Doi {10.1088/0034-4885/73/6/066901} {\bibfield  {journal}
  {\bibinfo  {journal} {Rept. Prog. Phys.}\ }\textbf {\bibinfo {volume}
  {73}},\ \bibinfo {pages} {066901} (\bibinfo {year} {2010})}\BibitemShut
  {NoStop}%
\bibitem [{\citenamefont {Koley}\ and\ \citenamefont
  {Kar}(2005){\natexlab{a}}}]{KoleyKar2005a}%
  \BibitemOpen
  \bibfield  {author} {\bibinfo {author} {\bibfnamefont {R.}~\bibnamefont
  {Koley}}\ and\ \bibinfo {author} {\bibfnamefont {S.}~\bibnamefont {Kar}},\
  }\Doi {10.1142/S0217732305015586} {\bibfield  {journal} {\bibinfo  {journal}
  {Mod. Phys. Lett. A}\ }\textbf {\bibinfo {volume} {20}},\ \bibinfo {pages}
  {363} (\bibinfo {year} {2005}{\natexlab{a}})}
  \BibitemShut {NoStop}%
\bibitem [{\citenamefont {Nunes}\ and\ \citenamefont
  {Peloso}(2005)}]{NunesPeloso2005}%
  \BibitemOpen
  \bibfield  {author} {\bibinfo {author} {\bibfnamefont {N.}~\bibnamefont
  {Nunes}}\ and\ \bibinfo {author} {\bibfnamefont {M.}~\bibnamefont {Peloso}},\
  }\Doi {10.1016/j.physletb.2005.07.047} {\bibfield  {journal} {\bibinfo
  {journal} {Phys. Lett. B}\ }\textbf {\bibinfo {volume} {623}},\ \bibinfo
  {pages} {147} (\bibinfo {year} {2005})}
  \BibitemShut {NoStop}%
\bibitem [{\citenamefont {Olechowski}(2008)}]{Olechowski2008}%
  \BibitemOpen
  \bibfield  {author} {\bibinfo {author} {\bibfnamefont {M.}~\bibnamefont
  {Olechowski}},\ }\Doi {10.1103/PhysRevD.78.084036} {\bibfield  {journal}
  {\bibinfo  {journal} {Phys. Rev. D}\ }\textbf {\bibinfo {volume} {78}},\
  \bibinfo {pages} {084036} (\bibinfo {year} {2008})} \BibitemShut
  {NoStop}%
\bibitem [{\citenamefont {Maity}\ \emph {et~al.}(2009)\citenamefont {Maity},
  \citenamefont {SenGupta},\ and\ \citenamefont {Sur}}]{MaitySenGuptaSur2009}%
  \BibitemOpen
  \bibfield  {author} {\bibinfo {author} {\bibfnamefont {D.}~\bibnamefont
  {Maity}}, \bibinfo {author} {\bibfnamefont {S.}~\bibnamefont {SenGupta}}, \
  and\ \bibinfo {author} {\bibfnamefont {S.}~\bibnamefont {Sur}},\ }\Doi
  {10.1088/0264-9381/26/5/055003} {\bibfield  {journal} {\bibinfo  {journal}
  {Classical Quantum Gravity},\ }\textbf {\bibinfo {volume} {26}},\ \bibinfo
  {pages} {055003} (\bibinfo {year} {2009})}
  \BibitemShut {NoStop}%
\bibitem [{\citenamefont {Goldberger}\ and\ \citenamefont
  {Wise}(1999)}]{GoldbergerWise1999}%
  \BibitemOpen
  \bibfield  {author} {\bibinfo {author} {\bibfnamefont {W.~D.}\ \bibnamefont
  {Goldberger}}\ and\ \bibinfo {author} {\bibfnamefont {M.~B.}\ \bibnamefont
  {Wise}},\ }\Doi {10.1103/PhysRevLett.83.4922} {\bibfield  {journal} {\bibinfo
   {journal} {Phys. Rev. Lett.}\ }\textbf {\bibinfo {volume} {83}},\ \bibinfo
  {pages} {4922} (\bibinfo {year} {1999})}\BibitemShut
  {NoStop}%
\bibitem [{\citenamefont {Lesgourgues}\ and\ \citenamefont
  {Sorbo}(2004)}]{LesgourguesSorbo2004}%
  \BibitemOpen
  \bibfield  {author} {\bibinfo {author} {\bibfnamefont {J.}~\bibnamefont
  {Lesgourgues}}\ and\ \bibinfo {author} {\bibfnamefont {L.}~\bibnamefont
  {Sorbo}},\ }\Doi {10.1103/PhysRevD.69.084010} {\bibfield  {journal} {\bibinfo
   {journal} {Phys. Rev. D}\ }\textbf {\bibinfo {volume} {69}},\ \bibinfo
  {pages} {084010} (\bibinfo {year} {2004})}\BibitemShut {NoStop}%
\bibitem [{\citenamefont {Adam}\ \emph
  {et~al.}(2008){\natexlab{a}}\citenamefont {Adam}, \citenamefont {Grandi},
  \citenamefont {Sanchez-Guillen},\ and\ \citenamefont
  {Wereszczynski}}]{AdamGrandiSanchez-GuillenWereszczynski2008}%
  \BibitemOpen
  \bibfield  {author} {\bibinfo {author} {\bibfnamefont {C.}~\bibnamefont
  {Adam}}, \bibinfo {author} {\bibfnamefont {N.}~\bibnamefont {Grandi}},
  \bibinfo {author} {\bibfnamefont {J.}~\bibnamefont {Sanchez-Guillen}}, \ and\
  \bibinfo {author} {\bibfnamefont {A.}~\bibnamefont {Wereszczynski}},\ }\Doi
  {10.1088/1751-8113/41/21/212004} {\bibfield  {journal} {\bibinfo  {journal}
  {J. Phys. A}\ }\textbf {\bibinfo {volume} {41}},\ \bibinfo {pages} {212004}
  (\bibinfo {year} {2008}{\natexlab{a}})} \BibitemShut
  {NoStop}%
\bibitem [{\citenamefont {Bazeia}\ \emph {et~al.}(2003)\citenamefont {Bazeia},
  \citenamefont {Brito},\ and\ \citenamefont
  {Nascimento}}]{BazeiaBritoNascimento2003}%
  \BibitemOpen
  \bibfield  {author} {\bibinfo {author} {\bibfnamefont {D.}~\bibnamefont
  {Bazeia}}, \bibinfo {author} {\bibfnamefont {F.~A.}\ \bibnamefont {Brito}}, \
  and\ \bibinfo {author} {\bibfnamefont {J.~R.}\ \bibnamefont {Nascimento}},\
  }\Doi {10.1103/PhysRevD.68.085007} {\bibfield  {journal} {\bibinfo  {journal}
  {Phys. Rev. D}\ }\textbf {\bibinfo {volume} {68}},\ \bibinfo {pages}
  {085007} (\bibinfo {year} {2003})}\BibitemShut {NoStop}%
\bibitem [{\citenamefont {Koley}\ and\ \citenamefont
  {Kar}(2005){\natexlab{b}}}]{KoleyKar2005b}%
  \BibitemOpen
  \bibfield  {author} {\bibinfo {author} {\bibfnamefont {R.}~\bibnamefont
  {Koley}}\ and\ \bibinfo {author} {\bibfnamefont {S.}~\bibnamefont {Kar}},\
  }\Doi {10.1016/j.physletb.2005.07.060} {\bibfield  {journal} {\bibinfo
  {journal} {Phys. Lett. B}\ }\textbf {\bibinfo {volume} {623}},\ \bibinfo
  {pages} {244} (\bibinfo {year} {2005}{\natexlab{b}})}
  \BibitemShut {NoStop}%
\bibitem [{\citenamefont {Adam}\ \emph
  {et~al.}(2008){\natexlab{b}}\citenamefont {Adam}, \citenamefont {Grandi},
  \citenamefont {Klimas}, \citenamefont {Sanchez-Guillen},\ and\ \citenamefont
  {Wereszczynski}}]{AdamGrandiKlimasSanchez-GuillenWereszczynski2008}%
  \BibitemOpen
  \bibfield  {author} {\bibinfo {author} {\bibfnamefont {C.}~\bibnamefont
  {Adam}}, \bibinfo {author} {\bibfnamefont {N.}~\bibnamefont {Grandi}},
  \bibinfo {author} {\bibfnamefont {P.}~\bibnamefont {Klimas}}, \bibinfo
  {author} {\bibfnamefont {J.}~\bibnamefont {Sanchez-Guillen}}, \ and\ \bibinfo
  {author} {\bibfnamefont {A.}~\bibnamefont {Wereszczynski}},\ }\Doi
  {10.1088/1751-8113/41/37/375401} {\bibfield  {journal} {\bibinfo  {journal}
  {J. Phys. A}\ }\textbf {\bibinfo {volume} {41}},\ \bibinfo {pages} {375401}
  (\bibinfo {year} {2008}{\natexlab{b}})}\BibitemShut
  {NoStop}%
\bibitem [{\citenamefont {Bazeia}\ \emph {et~al.}(2009)\citenamefont {Bazeia},
  \citenamefont {Gomes}, \citenamefont {Losano},\ and\ \citenamefont
  {Menezes}}]{BazeiaGomesLosanoMenezes2009}%
  \BibitemOpen
  \bibfield  {author} {\bibinfo {author} {\bibfnamefont {D.}~\bibnamefont
  {Bazeia}}, \bibinfo {author} {\bibfnamefont {A.~R.}\ \bibnamefont {Gomes}},
  \bibinfo {author} {\bibfnamefont {L.}~\bibnamefont {Losano}}, \ and\ \bibinfo
  {author} {\bibfnamefont {R.}~\bibnamefont {Menezes}},\ }\Doi
  {10.1016/j.physletb.2008.12.039} {\bibfield  {journal} {\bibinfo  {journal}
  {Phys. Lett. B}\ }\textbf {\bibinfo {volume} {671}},\ \bibinfo {pages} {402}
  (\bibinfo {year} {2009})}\BibitemShut {NoStop}%
\bibitem [{\citenamefont {Liu}\ \emph {et~al.}(2010)\citenamefont {Liu},
  \citenamefont {Zhong},\ and\ \citenamefont {Yang}}]{LiuZhongYang2010}%
  \BibitemOpen
  \bibfield  {author} {\bibinfo {author} {\bibfnamefont {Y.-X.}\ \bibnamefont
  {Liu}}, \bibinfo {author} {\bibfnamefont {Y.}~\bibnamefont {Zhong}}, \ and\
  \bibinfo {author} {\bibfnamefont {K.}~\bibnamefont {Yang}},\ }\Doi
  {10.1209/0295-5075/90/51001} {\bibfield  {journal} {\bibinfo  {journal}
  {Europhys. Lett.}\ }\textbf {\bibinfo {volume} {90}},\ \bibinfo {pages}
  {51001} (\bibinfo {year} {2010})} \BibitemShut {NoStop}%
\bibitem [{\citenamefont {Castro}\ and\ \citenamefont
  {Meza}(2013)}]{CastroMeza2013}%
  \BibitemOpen
  \bibfield  {author} {\bibinfo {author} {\bibfnamefont {L.~B.}\ \bibnamefont
  {Castro}}\ and\ \bibinfo {author} {\bibfnamefont {L.~A.}\ \bibnamefont
  {Meza}},\ }\Doi {10.1209/0295-5075/102/21001} {\bibfield  {journal} {\bibinfo
   {journal} {Europhys. Lett.}\ }\textbf {\bibinfo {volume} {102}},\ \bibinfo
  {pages} {21001} (\bibinfo {year} {2013})}\BibitemShut
  {NoStop}%
\bibitem [{\citenamefont {Bazeia}\ \emph {et~al.}(2011)\citenamefont {Bazeia},
  \citenamefont {Dantas}, \citenamefont {Gomes}, \citenamefont {Losano},\ and\
  \citenamefont {Menezes}}]{BazeiaDantasGomesLosanoMenezes2011}%
  \BibitemOpen
  \bibfield  {author} {\bibinfo {author} {\bibfnamefont {D.}~\bibnamefont
  {Bazeia}}, \bibinfo {author} {\bibfnamefont {J.~D.}\ \bibnamefont {Dantas}},
  \bibinfo {author} {\bibfnamefont {A.~R.}\ \bibnamefont {Gomes}}, \bibinfo
  {author} {\bibfnamefont {L.}~\bibnamefont {Losano}}, \ and\ \bibinfo {author}
  {\bibfnamefont {R.}~\bibnamefont {Menezes}},\ }\Doi
  {10.1103/PhysRevD.84.045010} {\bibfield  {journal} {\bibinfo  {journal}
  {Phys. Rev. D}\ }\textbf {\bibinfo {volume} {84}},\ \bibinfo {pages}
  {045010} (\bibinfo {year} {2011})}\BibitemShut {NoStop}%
\bibitem [{\citenamefont {Bazeia}\ \emph {et~al.}(2012)\citenamefont {Bazeia},
  \citenamefont {Lobao},\ and\ \citenamefont
  {Menezes}}]{BazeiaLobaoMenezes2012}%
  \BibitemOpen
  \bibfield  {author} {\bibinfo {author} {\bibfnamefont {D.}~\bibnamefont
  {Bazeia}}, \bibinfo {author} {\bibfnamefont {A. S.}~\bibnamefont {Lobao}}, \ and\ \bibinfo {author} {\bibfnamefont
  {R.}~\bibnamefont {Menezes}},\ }\Doi {10.1103/PhysRevD.86.125021} {\bibfield
  {journal} {\bibinfo  {journal} {Phys. Rev. D}\ }\textbf {\bibinfo {volume}
  {86}},\ \bibinfo {pages} {125021} (\bibinfo {year} {2012})} \BibitemShut
  {NoStop}%
\bibitem [{\citenamefont {Kobayashi}\ \emph {et~al.}(2002)\citenamefont
  {Kobayashi}, \citenamefont {Koyama},\ and\ \citenamefont
  {Soda}}]{KobayashiKoyamaSoda2002}%
  \BibitemOpen
  \bibfield  {author} {\bibinfo {author} {\bibfnamefont {S.}~\bibnamefont
  {Kobayashi}}, \bibinfo {author} {\bibfnamefont {K.}~\bibnamefont {Koyama}}, \
  and\ \bibinfo {author} {\bibfnamefont {J.}~\bibnamefont {Soda}},\ }\Doi
  {10.1103/PhysRevD.65.064014} {\bibfield  {journal} {\bibinfo  {journal}
  {Phys. Rev. D}\ }\textbf {\bibinfo {volume} {65}},\ \bibinfo {pages}
  {064014} (\bibinfo {year} {2002})} \BibitemShut
  {NoStop}%
\bibitem [{\citenamefont {Weinberg}(2008)}]{Weinberg2008}%
  \BibitemOpen
  \bibfield  {author} {\bibinfo {author} {\bibfnamefont {S.}~\bibnamefont
  {Weinberg}},\ }\href@noop {} {\emph {\bibinfo {title} {Cosmology}}}\
  (\bibinfo  {publisher} {Oxford University Press},\ \bibinfo {year}
  {2008})\BibitemShut {NoStop}%
\bibitem [{\citenamefont {Mukhanov}(2005)}]{Mukhanov2005}%
  \BibitemOpen
  \bibfield  {author} {\bibinfo {author} {\bibfnamefont {V.}~\bibnamefont
  {Mukhanov}},\ }\href@noop {} {\emph {\bibinfo {title} {Physical Foundations
  Of Cosmology}}}\ (\bibinfo  {publisher} {Cambridge University Press},\
  \bibinfo {year} {2005})\BibitemShut {NoStop}%
\bibitem [{\citenamefont {Kakushadze}\ and\ \citenamefont
  {Langfelder}(2000)}]{KakushadzeLangfelder2000}%
  \BibitemOpen
  \bibfield  {author} {\bibinfo {author} {\bibfnamefont {Z.}~\bibnamefont
  {Kakushadze}}\ and\ \bibinfo {author} {\bibfnamefont {P.}~\bibnamefont
  {Langfelder}},\ }\Doi {10.1142/S0217732300002693} {\bibfield  {journal}
  {\bibinfo  {journal} {Mod. Phys. Lett. A}\ }\textbf {\bibinfo {volume}
  {15}},\ \bibinfo {pages} {2265} (\bibinfo {year} {2000})}
  \BibitemShut {NoStop}%
\bibitem [{\citenamefont {Randjbar-Daemi}\ and\ \citenamefont
  {Shaposhnikov}(2002)}]{Randjbar-DaemiShaposhnikov2002}%
  \BibitemOpen
  \bibfield  {author} {\bibinfo {author} {\bibfnamefont {S.}~\bibnamefont
  {Randjbar-Daemi}}\ and\ \bibinfo {author} {\bibfnamefont {M.}~\bibnamefont
  {Shaposhnikov}},\ }\Doi {10.1016/S0550-3213(02)00828-3} {\bibfield  {journal}
  {\bibinfo  {journal} {Nucl. Phys. B}\ }\textbf {\bibinfo {volume} {645}},\
  \bibinfo {pages} {188} (\bibinfo {year} {2002})}
  \BibitemShut {NoStop}%
\bibitem [{\citenamefont {Mukhanov}\ \emph {et~al.}(1992)\citenamefont
  {Mukhanov}, \citenamefont {Feldman},\ and\ \citenamefont
  {Brandenberger}}]{MukhanovFeldmanBrandenberger1992}%
  \BibitemOpen
  \bibfield  {author} {\bibinfo {author} {\bibfnamefont {V.~F.}\ \bibnamefont
  {Mukhanov}}, \bibinfo {author} {\bibfnamefont {H.~A.}\ \bibnamefont
  {Feldman}}, \ and\ \bibinfo {author} {\bibfnamefont {R.~H.}\ \bibnamefont
  {Brandenberger}},\ }\Doi {DOI: 10.1016/0370-1573(92)90044-Z} {\bibfield
  {journal} {\bibinfo  {journal} {Phys. Rep.}\ }\textbf {\bibinfo {volume}
  {215}},\ \bibinfo {pages} {203} (\bibinfo {year} {1992})}\BibitemShut
  {NoStop}%
\bibitem [{\citenamefont {Gong}\ \emph {et~al.}(2010)\citenamefont {Gong},
  \citenamefont {Koh},\ and\ \citenamefont {Sasaki}}]{GongKohSasaki2010}%
  \BibitemOpen
  \bibfield  {author} {\bibinfo {author} {\bibfnamefont {J.-O.}\ \bibnamefont
  {Gong}}, \bibinfo {author} {\bibfnamefont {S.}~\bibnamefont {Koh}}, \ and\
  \bibinfo {author} {\bibfnamefont {M.}~\bibnamefont {Sasaki}},\ }\Doi
  {10.1103/PhysRevD.81.084053} {\bibfield  {journal} {\bibinfo  {journal}
  {Phys. Rev. D}\ }\textbf {\bibinfo {volume} {81}},\ \bibinfo {pages}
  {084053} (\bibinfo {year} {2010})}\BibitemShut {NoStop}%
\end{thebibliography}

\end{document}